\documentstyle[12pt,epsf]{article}
\begin{document}
\textwidth=14cm
\textheight=21cm
\def\ds{\displaystyle}

\centerline{\bf GROUND STATE OF QUANTUM JAHN-TELLER MODEL:}
\centerline{\bf SELFTRAPPING VS CORRELATED PHONON-ASSISTED} 
\centerline{\bf TUNNELING }

\vspace{1cm}

\centerline{ Eva Majern\'{\i}kov\'a${}^{\dag \ddag }$,
and S. Shpyrko${}^{\dag }$ }
\centerline{${}^{\dag}$Department of Theoretical Physics, Palack\'y University,}
\centerline{T\v r. 17. listopadu 50, CZ-77207 Olomouc, Czech Republic}
\centerline{${}^{\ddag}$Institute of Physics, Slovak Academy of Sciences,
D\'ubravsk\'a cesta,}\centerline{ SK-84 228 Bratislava, Slovak Republic
}

\begin{abstract}

Ground state of the quantum Jahn-Teller model with broken rotational
symmetry was investigated by the variational approach in two cases:
a lattice and a local ones.
Both cases differ by the way of accounting for the nonlinearity hidden
in the reflection-symmetric Hamiltonian. In spite of that the ground
state energy in both cases shows the same features:
There appear two regions of model parameters governing the
ground state: the region of dominating selftrapping
modified by the quantum effects and
the region of dominating phonon-assisted tunneling (antiselftrapping).
In the local case (i) the effect of quantum fluctuations and anharmonicity due
to the two-mode correlations is up to two orders larger than contributions
 due to the reflection effects of two-center wave function;
(ii) the variational results for the ground state energy
were compared with exact numerical results. The
coincidence is the better the more far away from the transition region
at the E$\otimes$e symmetry where the variational approach fails.
\end{abstract}

\centerline{\bf List of Content}

\begin{enumerate}
\item Introduction
\item Extended (lattice) generalized Jahn-Teller model
\item  Ground state of the lattice model
\item Ground state of the local model
\item Discussion of the numerical results
\item Quantum fluctuations in the local model
\item Conclusion
\end{enumerate}



\section{Introduction}

 Jahn-Teller model as a protopype model for phonon
removal of degeneracy of electron levels in complex molecules
\cite{Obrien:1993}  was investigated  mainly in its rotational symmetric
E$\otimes$e local version with electron coupling to two degenerate
intramolecular phonon modes, one antisymmetric
and one symmetric
against reflection.
Importance of focusing on the lattice version of the model increased due to
JT-based structural phase transitions in some recently discovered
high-$T_c$ superconductors and manganese-based
perovskites \cite{Kaplan:1995}, \cite{Gunnarson:1995}.

For the reflection symmetric two-level electron-phonon models with
linear coupling to one phonon mode (exciton, dimer)
 Shore et al. \cite{Sander:1973} introduced
variational wave function in a form of linear combination of the
harmonic oscillator wave functions related with two levels.
Two asymmetric minima of effective polaron potential turn coupled
by a variational parameter respecting its anharmonism by assuming
two-center variational phonon wave function.
This approach was shown to yield the
lowest ground state energy for the two-level models
\cite{Sander:1973}, \cite{Wagner:1994}.

The peculiarities due to reflection phenomena are also expected in the
case of linear coupling with two phonon modes.
In contrast to the model with coupling to one
phonon mode, the oscillations mediated by the antisymmetric mode are
nonlinearly coupled to the symmetric phonon mode.
Therefore, in order to improve the harmonic oscillator variational
ansatz for this model,
it was necessary to introduce additional variational parameters (VP)
considering {\it correlation}
 of both phonon modes. This was performed by Lo
\cite{Lo:1991a} for the E$\otimes$e JT
model and by Lo et al. \cite{Lo:1991b}) for the dimer model.
 The correlation was found most effective in the region of parameters
where the competing localization (polaron)
energy and the delocalization transfer energy (tunneling) were comparable.

In crystals exhibiting high space anisotropy the rotational symmetry
of Jahn-Teller molecules can be broken.
Therefore it is reasonable to investigate JT model generalized by breaking the
rotational symmetry (while saving the reflection symmetry): we assume
different coupling strengths for the onsite (intralevel) and interlevel
electron-phonon couplings.
Such a model can also be considered as a generalization of the
exciton-phonon or the dimer-phonon model by assuming the electron tunneling
 phonon assisted.
 JT model with the broken rotational symmetry will
involve the coupling of two minima as well as
 the coupling of the symmetric and antisymmetric phonon modes.
In the lattice model, however, in contrast to the local case,
the mode correlation appears to be of a marginal importance (Section
III).
 As a consequence, our ansatz for the variational wave function of the
local model will involve (i) the reflection
VP introduced by the construction of the two-center phonon wave
function, and
 (ii) correlation VP respecting coupling of the symmetric and antisymmetric
 phonon modes.
 The question arises of the importance of these VPs and respective
 effects regarding their relevance in different regions of the model
 parameters.
 Formulation of the variational ansatz of the local case
 is presented in the Section IV.
 Calculation of the ground state energy of the lattice model
 and related discussions are the contents of the Sect.III. and V.
Reliability of different variational alternatives was checked by
comparison with results of exact numerical simulations for the local
model.

\section{Extended (lattice) generalized Jahn-Teller model }

We investigate 1D lattice of spinless double degenerated electron
states linearly coupled to two intramolecular phonon modes
described by Hamiltonian

 \begin{eqnarray}
H= \Omega \sum_{n,i=1,2} (b_{i,n}^{\dag}b_{i,n} +1/2
 ) +\sum\limits_{n}
\left(  \alpha  (b_{1,n}^{\dag}+b_{1,n})\sigma_{zn}
\right.\nonumber\\
\left. -\beta (b_{2,n}^{\dag}+b_{2,n})\sigma_{xn}\right )
-\frac{T}{2} \sum_{n,j=1,2} (R_{1,j}+R_{-1,j})I_{n} \, .
\label{Hams1}
\end{eqnarray}
where $b_{i,n}, i=1,2$ are phonon annihilation operators, and
 the Pauli matrices $\sigma_{ln}$ represent two-level electron system.
 They satisfy identities
$\ [\sigma_{ln},\sigma_{jn}]=i\sigma_{kn},
 \ l=x,y,z$, representing $1/2$-pseudo-spins related to
 the electron densities in a usual way, i.e.
$\sigma_{xn}=\frac{1}{2}(c^{\dag}_{1,n}c_{2,n}+
c^{\dag}_{2,n}c_{1,n}),\ $
$\sigma_{yn}=\frac{1}{2i}(c^{\dag}_{1,n}c_{2,n}-
c^{\dag}_{2,n}c_{1,n}),  $\quad
$\sigma_{zn}=\frac{1}{2}(c^{\dag}_{1,n}c_{1,n}-
c^{\dag}_{2,n}c_{2,n}),$ $ \  I_n= \frac{1}{2} (c^{\dag}_{1,n}c_{1,n}+
c^{\dag}_{2,n}c_{2,n}) $ is a unit matrix, and
$c_{j,n}$ are electron annihilation operators.
The operator $R_{\pm 1,j}=e^{\pm ipa}$ of the displacement by a
lattice constant $\pm a$ acts in both the electron and phonon space,
$R_{\pm 1,j}f_n=f_{n\pm 1} R_{\pm 1,j} $.

In terms of the creation-annihilation electron and phonon operators
the Hamiltonian can be cast as follows:

 \begin{eqnarray}
H=\sum_{n} [\Omega\sum_{i=1,2} (b_{i n}^{\dag}b_{i n}
+\frac{1}{2} ) +\frac{\alpha}{2}(c^{\dag}_{1,n}c_{1,n}-
c^{\dag}_{2,n}c_{2,n})(b_{1n}^{\dag}+b_{1n})
\nonumber\\
-\frac{\beta}{2} (c_{1n}^{\dag}c_{2n}+c_{2n}^{\dag}c_{1n}
)(b_{2n}^{\dag}+b_{2n})-\frac{T}{2} \sum_{j=1,2}( 
c_{j,n}^{\dag}c_{j,n+1}+H.c. )]\, .\qquad
\label{1}
\end{eqnarray}

For  $\beta=-\alpha $, the  interaction part of (\ref{Hams1})
\begin{equation}
\alpha \left (\matrix{ b_{1n}^{\dag}+b_{1n}, \ b_{2n}^{\dag}+ b_{2n}\cr
 b_{2n}^{\dag}+b_{2n} ,\  -( b_{1n}^{\dag}+b_{1n})   }\right)
 \label{sym}
\end{equation}
yields the rotationally symmetric $E\otimes e$ form  \cite{Obrien:1993}
with a pair (an antisymmetric and a symmetric under reflection) of
double degenerated vibrations. This is, e.g., the case of $Cu^{++}$
ions with $d^9$ configurations in high-$T_c$
cuprates \cite{Obrien:1993},\cite{Kaplan:1995}.

 Taking $\alpha\neq \beta $ removes the
 degeneration of the vibronic states breaking the
 rotational symmetry of the electron-phonon interactions, the
 model still staying within the class
 of JT models \cite{Obrien:1993}, \cite{Kaplan:1995}.

  The dispersionless optical phonon mode $b_{1}$ splits
the degenerated unperturbed electron level ($j=1,2$)
while the mode $b_{2}$ mediates the electron
transitions between the levels. This latter term represents phonon-assisted
tunneling, a mechanism of the nonclassical (nonadiabatic) nature as well as
 is the pure tunneling in related exciton and dimer models.

 Evidently, Hamiltonian (\ref{Hams1}) ($\alpha\neq \beta$)
 is reflection-symmetric,  $G^{(el)} G_1^{(ph)}H=H $,
  \begin{eqnarray}
|2\rangle &=& G^{(el)} |1\rangle, \ (G^{(el)})^ 2=1,\nonumber\\
 G^{(ph)}_{1n} (b_{1n}^{\dag}\pm b_{1n}) &=&
- (b_{1n}^{\dag}\pm b_{1n}) G^{(ph)}_{1n}, \  (G^{(ph)}_{1n})^ 2=1,
\label{Gn}
\end{eqnarray}
where 
$G^{(ph)}_{1n}= \exp (i\pi b^{\dag}_{1n}b_{1n})$ is the phonon
reflection operator. While the phonon $1$ is antisymmetric under the
reflection, phonon $2$ remains symmetric.

In addition, the transfer part of (\ref{1}) exhibits $SU(2)$ symmetry
of the left- and right-moving electrons (holes).

Let us note that the quantum phonon assistance of the
electron tunneling ($\beta$-term in (\ref{1}) and (\ref{Hams1}))
constitutes the difference of the model from
the related dimer and exciton quantum models where instead of $\beta
\sum \limits _n (b_{2n}^{\dag}+b_{2n})\sigma_{xn}$ of (\ref{Hams1})
there stands $\Delta\sigma_{xn}$, where $\Delta $
is the distance between the levels
\cite{Sander:1973},\cite{Wagner:1994}.

The local part of (\ref{Hams1}) can be diagonalized
in electron subspace by the Fulton-Gouterman  unitary operator
\cite{Fulton:1961} $U_n\equiv U_{2,n}U_{1,n}$, where
\begin{equation}
 U_{i,n}= \frac{1}{\sqrt 2} \left ( \matrix{1\ , \ G_{i,n}\cr  1\ ,
\ -G_{i,n}}\right ), \quad G_{i,n} =\exp (i\pi
b_{in}^{\dag}b_{in})\equiv G^{(ph)}_{i,n},
\label{Un}
\end{equation}
 as  follows
 \begin{eqnarray}
\tilde H_L
=\sum\limits_n  U_n H_{0n}U_n^{-1}=
\Omega\sum\limits_{n, i=1,2}
\left (b_{in}^{\dag}b_{in} +\frac{1}{2}\right )\nonumber\\
+\frac{1}{2}\sum\limits_{n}[\alpha (b_{1n}^{\dag}+b_{1,n})
-\beta (b_{2n}^{\dag}+b_{2n})G_{1,n})]I_n .
\label{diagh1}
\end{eqnarray}
On the other hand, in the transfer term
\begin{equation}
\tilde H_T = -\frac{T}{2} \sum_{n}\left (V_{n,1}R_1 + V_{n, -1}R_{-1}\right )
\label{diagh2}
\end{equation}
there appears a nondiagonality
\begin{eqnarray}
V_{n,\pm 1}= \left [(1+G_{1n}G_{1n \pm 1})I_n+
(1-G_{1n}G_{1,n\pm 1})G_{2n}\sigma_{zn}\right ]\nonumber\\
\times\left [(1+G_{2n}G_{2n \pm 1})I_n+ (1-G_{2n}G_{2,n\pm 1})
\sigma_{xn}\right ]. \qquad
\label{V}
\end{eqnarray}
Here, Pauli matrices transform as 
$U_{i} \sigma_{x}U_{i}^{-1}=  G_{i}\sigma_z$, $U_{i} \sigma_z U_{i}^{-1}=
\sigma_x$, and $ U_{i} (b_i^{\dag}+ b_i) U^{-1}_i = (b_i^{\dag}+b_i) 2\sigma_x$.

The diagonal terms of (\ref{V}) represent the polaron transfer within
one level while the off-diagonal ones represent the interlevel polaron transfer
through the lattice. Evidently, the
contribution of the off-diagonal terms proportional to $1-G_{in}G_{i,n+1}$ is
much smaller when compared with those proportional to
$1+G_{in} G_{i,n+1}$.

Because of nonconservation of the number of
coherent phonons, they are able even in
the ground state to assist electron transitions between the levels.
In the Hamiltonian (\ref{diagh1}),
the operator  $G_{1n} =(-1)^{b_{1n}^{\dag}b_{1n}}$ (\ref{Un}),
highly nonlinear in the phonon-$1$ appears mediated by phonons $2$.
It introduces multiple electron oscillations between the
split levels mediated by {\it continuous virtual absorption and emission of the
 phonons $1$}.
The effect is analogous  to Rabi oscillations in quantum optics
due to photons \cite{Rabi:1936}.
Let us note that Rabi oscillations assist  both the interlevel
onsite and intersite electron transitions mediated by the
electron transfer $T$.

Further we shall investigate by variational approach the two variants 
of the model (\ref{Hams1}) - the lattice and the local one 
(considering $T=0$).

Shore et al \cite{Sander:1973} and Wagner et al
\cite{Wagner:1994}  proved for exciton or dimer models coupled to one
phonon mode that the two-center wave function generalized to an
asymmetric nonunitary ansatz with a variational parameter $\eta$ of a form

\begin{equation}
\Psi ^{(p)}=  \frac{1}{\sqrt  C}(1+\eta p G)|1\rangle \phi^{(p)},
\label{6}
\end{equation}
(where $C$ is a normalization constant), yields
lower ground state \cite{Sander:1973}, \cite{Wagner:1994}.

So, rather generally we
define phonon wave functions $\phi^{(p)}$
\begin{equation}
\phi^{(\pm)}=  D_1(\pm \zeta_1) D_2(\zeta_2)S_2(r_2)S_1(r_1)S_{12}
(\pm\lambda)|0\rangle.
\label{7}
\end{equation}

Here, the generators of variational displacements  are defined
\begin{equation}
D_i (\eta)= \exp [\zeta_i(b_i^{\dag}-b_i)],
\label{8}
\end{equation}
and those of squeezing
\begin{equation}
S_i(r_i)= \exp[r_i(b_i^{\dag 2}-b_i^{2})],
\label{9}
\end{equation}
for $i=1,2$. In Eq. (\ref{diagh1}) coupling of the phonon modes $1$ and
$2$ occurs, therefore one includes into the ansatz (\ref{7}) also
the correlation generator
\begin{equation}
 S_{12}(\lambda)= \exp[\lambda (b_1^{\dag}b_2^{\dag}-b_1b_2)].
 \label{10}
 \end{equation}
 with a correlation VP $\lambda$.

The form (\ref{7}) is written here seemingly for the local case, while the 
most general form of it should contain the dependencies of all variational 
parameters on $n$ (site number in the lattice): $\gamma_1 \to \gamma_{1m}(n)$
etc, showing the displacement of the mode $1$ at the site $m$ due to the 
electron at the site $n$, $\gamma_{jm}(n)=
\frac{1}{\sqrt N}\sum \limits_q \gamma_{jq} (n)\exp (-iqma)$.
We take
\begin{equation}
\gamma_{j q}(n)= \frac{\gamma_{j}}{\sqrt N} \exp(iqna)\rightarrow
\gamma_{j m}(n)=\gamma_j \delta_{m,n},
\label{Q}
\end{equation}
where $\gamma_j $ are independent of $n$. Eq. (\ref{Q}) indicates
that the phonon displacement accompanies the electron at the site $n$.
The same is valid for the squeezing, $r_m (n)=
r\delta_{m,n}$ and mixing parameter $\lambda$. When we consider the ground 
state of the lattice model, we omit the electron and phonon dynamic terms by 
setting $k=0$, $q=0$, thus even in the lattice case we are left with the 
effective local molecule Hamiltonian, the only difference being the additional 
contribution of the transfer terms (those containing $T$). As it is shown later,
in the lattice model we can set $\lambda=0$ without worsening considerably
the variational results, while for the local case $T=0$ introducing
this parameter considerably improves the results. In what follows the 
ground state energy as function of optimized
variational parameters $\eta$,
$\gamma_{1} ,\gamma_{2}, r, \lambda $ will be determined.

  \section{Ground state of the lattice model}

First we consider the lattice model setting $\lambda=0$. 

By averaging $\tilde H= \tilde H_L+ \tilde H_{T}$ ((\ref{diagh1}),
(\ref{diagh2})) over the  phonon wave functions
(\ref{6}) with (\ref{7}) one obtains for the site
Hamiltonian (\ref{diagh1}) the local part in the form
 \begin{equation}
\frac{\langle \tilde H_L\rangle}{N}\equiv H_{ph} + H_{int} ,
\label{HD}
 \end{equation}
 where
\begin{eqnarray}
& &  H_{ph} = \frac{\Omega}{2}(\cosh 4r +1) I
+ \Omega \frac{1+\eta^2-2\eta e^{-8r_1}\exp(-2|\tilde \gamma_{1}|^2)}
{1+\eta^2+2\eta\exp(-2|\tilde \gamma_{1}|^2)} |\gamma_{1}|^2 I
+ \nonumber \\
& & +
\Omega |\gamma_{2 }|^2 I,
\label{hav2}
\end{eqnarray}

\begin{eqnarray}
 & & H_{int} \equiv 
 H_{\alpha}+H_{\beta} 
= \frac{\alpha (1-\eta^2)}
 {1+\eta^2+2\eta\exp(-2|\tilde \gamma_{1}|^2)}\gamma_{1}\sigma_{z}
-  \nonumber \\ & & -
\frac{\beta [(1+\eta^2)\exp(-2|\tilde \gamma_{1q}|^2)+2\eta]}
{1+\eta^2+2\eta\exp(-2|\tilde \gamma_{1}|^2)}
\gamma_{2} I \, ,
\label{hav12}
\end{eqnarray}
where $H_{\alpha}$ and $H_{\beta}$ are $\alpha$- and $\beta$-dependent parts
of the interaction $H_{int}$ (\ref{hav12}).
From the transfer Hamiltonian $\tilde H_T$ (\ref{diagh2}) there remains
\begin{eqnarray}
\frac{\langle\tilde H_T\rangle }{N}=-T  \exp(-W )M
\equiv 2(H_{T1}I+H_{T2}i \sigma_{y}+ H_{T3}\sigma_{z} + H_{T4}\sigma_x
).
\label{havt}
\end{eqnarray}
where $M$:               
\begin{eqnarray}
M = \left [( 1+E_{1}^2 )I +
(1-E_1^2) E_2 \sigma_{zn} \right ]
. \left [(1+E_2^2) I + (1-E_2^2)\sigma_{x} \right ],
\label{T2}
\end{eqnarray}
and  $E_i$ are given by 
\begin{equation}
  E_1 \equiv \exp\left (-2|\tilde\gamma_{1}|^2\right ), \
   E_2 \equiv \exp\left (-2|\gamma_{2}|^2\right ).
   \label{Ei}
\end{equation}
Expressions $E_i$ (\ref{Ei}) in the ground state are independent of
$q$ and $n$ because of the form of $\gamma_{iq}(n)=
\gamma_i \exp (-iqna)$ (\ref{Q}).
This substantially simplifies subsequent calculations leaving
$E_{i}$ and $ C_{1}$ as functions of $\gamma_i^2 $ independent on $n$.
The transfer terms $H_{Ti}, \
i=1,2,3,4$ in (\ref{havt}) with (\ref{T2}) are expressed as
\begin{eqnarray}
H_{T1}= -\frac{T}{4} (1+E_1^2)(1+E_2^2),
\quad H_{T2}= -\frac{T}{4}(1-E_1^2)(1-E_2^2)E_2,
\nonumber\\
H_{T3}= -\frac{T}{4} (1-E_1^2)(1+E_2^2)E_2,\quad
 H_{T4}= -\frac{T}{4} (1+E_1^2)(1-E_2^2).
\label{hti}
\end{eqnarray}
  Here, $H_{T1}, H_{T3}$ and $H_{T2}, H_{T4}$  are
diagonal and off-diagonal terms of (\ref{havt}), respectively.


The effective polaron potential in (\ref{HD}) (with
 (\ref{hav2})-(\ref{Ei})) as a highly nonlinear function of
 $\gamma_1$ and $\gamma_2$ is visualized in Fig. 1.
 \begin{figure}[h]
\epsfxsize=9cm
\epsfbox{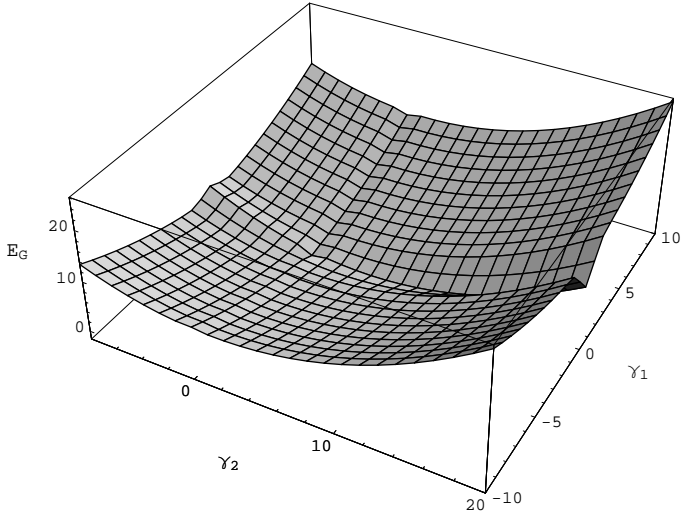}
\caption{ The effective potential for $\chi=\beta/\alpha=0.5, \mu=2$ and
 $T=1, \Omega=0.05$ as function of $\gamma_1$, $\gamma_2$.  }
\label{fig1}
\end{figure}

The potential exhibits two sets of minima related to two competing
ground states:\\
(i)
Two nonequivalent broad minima related to both the levels
(\ref{7}) at $\pm \gamma_1 \neq 0$ and $\gamma_2$ close to $0$; \\
(ii) One narrow minimum at $\gamma_1$ close to $0$, $\gamma_2 >0$, where
both levels approach close together. This minimum develops at 
growing $\beta$; evidently its behaviour depends also on $T$ value
because of nonlinearity of the Debye-Waller factor $M$ ((\ref{T2}) and
(\ref{Ei})).

The ground state energies related to these two sets of competing minima
were calculated numerically.
The result of the numerical minimalization of the diagonalized form of energy
(\ref{HD})
\begin{eqnarray}
E=  H_{ph}+H_{\beta }+ H_{T_1}-\left [(H_{\alpha}+H_{T_3})^2+
H^2_{T_4}-H^2_{T_2}\right ]^{1/2}
\label{EF}
\end{eqnarray}

 can be written symbolically as
\begin{equation}
E_G (\mu, \chi,\Omega/T)= E (\{\gamma_{1q}, \gamma_{2q}, r, \eta
\}_{min})/T.
\label{HG}
\end{equation}
Here, the index $min$ denotes the optimized values.
The model parameters
\begin{equation}
\mu=\frac{\alpha^2}{2\Omega T},\quad
\chi=\frac{\beta}{\alpha}, \quad \frac{\Omega}{T}
\label{mu}
\end{equation}
are parameters of the effective
interaction, asymmetry and nonadiabaticity, respectively. Energy $E_G$
(\ref{HG}) is in scaled units, renormalized by $T$. The results of the
numerical evaluations of the ground state energy (\ref{HG}) are depicted
in Figs. 2a, 2b  different model parameters.

 One can distinguish there two regions depending on the value of $\chi$ with different behaviour of ground state in each of them:\\
(i) The ground state pertaining to the lower broad minimum at $\gamma_1 <0$
 and $\gamma_2\approx 0$ with a small reflection part at $\gamma_1>0$
 is referred to as a "heavy" region.
It corresponds to predominantly intralevel "heavy" polaron which is
represented by a two-peak wave function,
both peaks representing a harmonic oscillator ($\sim \exp
[-(x\mp \gamma_1)^2]$) displaced by the value $\pm \gamma_1$ (as it is seen
from the form of the ansatz (\ref{7})). The "heavy" polaron
is dominant at $\chi<1$ where the broad minimum  $\gamma_1<0$ is dominating.\\
(ii) Close to $\chi=1$, the
energies of two minima go very close together (their difference is
of the order of the phonon energy $\Omega$). They
drop to one narrow minimum which represents a new ground state.
{\it Close to $\chi=1$, continuous transition to a new ground state occurs}.
It stabilizes at $\chi>1$, where (optimized) $\gamma_1$ value is close to $0$ and $\gamma_2>0$.
 This region is referred to as a "light" polaron region because, owing
 to the abrupt decrease of $\gamma_1\approx 0$ at $\chi>1$, the
 effective mass of the intralevel polaron drops to almost its
 free-electron value, i.e., {\it the polaron selflocalization vanishes
 in the "light" region}.
 Because of this "undressing", the transport characteristics of the excited
 electron would increase. Moreover, due to the tiny distance between the
 levels, their coupling takes place by the exchange of virtual phonons $1$.
  At suitable conditions, in the excited state, when both
 levels are occupied by electrons of opposite spins,
 {\it the mechanism of virtual phonon exchange implies
 the pairing of electrons, i.e. formation of "light" bipolarons }.

 \begin{figure}[h]
\epsfxsize=12cm
\epsfbox{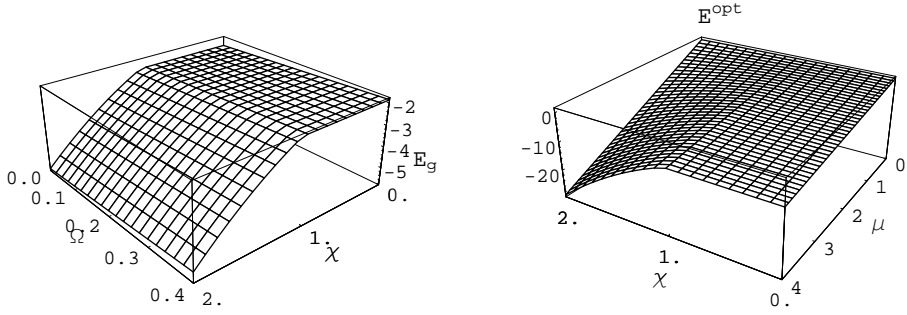}
\caption{The ground state energy  (\ref{11}) in the $\chi $-$\Omega $
plane at $\mu=2.5$ (a) and in the plane $\chi - \mu $ ($\Omega=1$)  (b).
The seltrapping dominated phase at $\chi<1$ and the tunneling
dominated phase at $\chi>1$.}
\label{fig2}
\end{figure}

 The ground state energy,
 especially its behavior dependent on pairs of parameters $\chi$,
 $\Omega$ and  $\chi$, $\mu$ is illustrated in details in
 Fig. 2. While being weakly $\Omega$-dependent, the energy  strongly
decreases with $\chi$ inside the "light" phase (Fig. 2a).
 The position of the phase line is slightly shifted from $\chi=1$ at
$\Omega=0$ to higher values of
$\chi$ with increasing $\Omega$.  This is consistent with
 the fact that the phonon fluctuations are most effective
 when the difference of the energies of the phases is of the order
 of the phonon energy.

The ground state energy in the "heavy" region ($\chi<1$) is
 independent of $\chi$, its decrease inside the "light" region
($\chi>1$) is dependent on the effective coupling $\mu$ (Fig. 2b).
The energy decrease due to $\mu$ is stronger in the "light" phase
(depending on $\chi$) than in the "heavy" phase.

 \begin{figure}[h]
\epsfxsize=10cm
\epsfbox{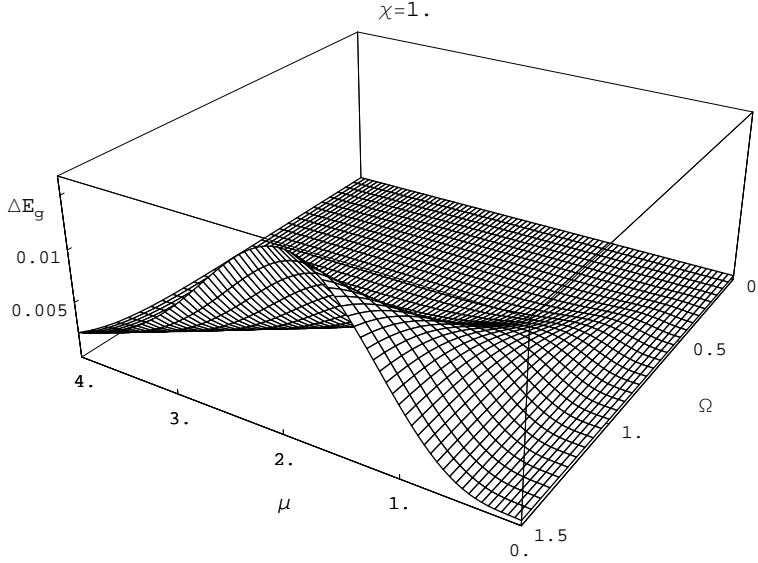}
\caption{The range of relevance of the reflection parameter $\eta$:
Difference  $\Delta E_G=E_G(\eta=0, r=0)-E_G$, for $\chi=1$,
 $E_G $ is the exact ground state (\ref{HG}). For $\chi<1$, the
difference increases with $\chi$ reaching its maximum at $\chi=1$.
At $\chi>1$ it drops to zero,
i.e. the narrow minimum "light" phase is resistant against $\eta$. }
\label{fig3}
\end{figure}

 The region of importance of reflection measure $\eta $  and $r$ is
 illustrated in the
 Fig. 3. There we show the difference between the ground state energy
 with $r$ and $\eta$ omitted and ground state with four variational
 parameters calculated numerically. However,  contribution due to $r$
 $E_{G}(\eta, r=0)-E_G \sim 2.10^{-3}$ of the maximum value in Fig. 3.
 The said parameters turn out to be most relevant in E$\otimes$e JT
 case ($\chi=1$).

The effects of $\eta $ and $r$ on the ground state is apparent
only at moderate couplings in the "heavy" region,  reaching their
maximum at $\chi=1$.
 The narrow minimum (the "light" region) is resistant against $\eta$.

In order to demonstrate important properties of displacements ($\mu $
and $\Omega$ dependence) it is sufficient, except of the region of
importance of the reflection (Fig. 3) inside the "heavy" region
apparent for $\gamma_2$,
to calculate them in the limit $\eta=0$.
We shall use Eq. (\ref{EF}) approximated for the cases of both "heavy" and
"light" polaron and obtain implicit expressions for $\gamma_i$ which however
are good to visualize their behaviour in both regions:

(i) "heavy" polaron ($\gamma_2$ small, $ E_2\approx 1,  H_{T_2},
H_{T_4}\approx 0$):
\begin{eqnarray}
\gamma_{1} & \approx &\frac{-\alpha e^{4r} }
{2T E_1^2\left
[2+\frac{\Omega e^{4r}}{T E_1^2} \right ]},
\label{ie1}
\end{eqnarray}
$\gamma_2 $ is small except of the region of fluctuations
visualized in Fig. 4. One can see, that
the selflocalization due to phonons $2 $ in the "heavy" region
$\sim - \gamma_2^2$ clearly implies the "cave"  in the ground
state energy due to the reflection measure $\eta$ (anharmonicity) of the
ground state (Fig. 3).

(ii) "light" polaron ($\gamma_1$ small, $ E_1\approx 1,
H_{T_2}, H_{T_3}\approx 0$):
\begin{eqnarray}
\gamma_{1}& \approx & \frac{-\alpha  e^{4r} }{2 T E_1^2
\left (1+\frac{\beta^2 }{2\Omega T}+  \frac{\Omega e^{4r}}{TE_1^2}\right
)}\approx 10^{-8},   \quad
 \gamma_{2} \approx \frac{\beta}
{2TE_2^2(2+\frac{\Omega}{TE_2^2})}.
\label{ie2}
\end{eqnarray}
In the "light" region the fluctuations of $\gamma_1$ are missing as well as
the fluctuations of the energy. This is consistent with the above result
of the resistance of the narrow minimum against $\eta$.

For both "heavy" and  "light" polaron a dependence of
$\gamma_{i}$ on the nonadiabaticity parameter $\Omega/T$ appears.
It implies the dependence of
the Debye-Waller factor and consequently of the polaron mass
on the phonon frequency $\Omega$. This can be thought of as an analogy
of the {\it isotope effect} at zero temperature.

In the "heavy" phase, electron transitions mediated by phonons $2$ to the
upper level enhance fluctuations $\sim \gamma_2$ which mix phonons $2$ with
phonons $1$ and contribute to the fluctuations of the ground state
energy of the heavy region. This is the reason for the similarity of
the results in Fig. 3.

\section{Ground state of the local model}

Now we consider the same Hamiltonian, but taken for the local molecule
($T=0$). From the outset we include the additional variational mixing 
parameter $\lambda$ into the Ansatz. The understanding of the importance of
introducing the mixing parameter in the local model in the contrast to the
lattice one can be gained if one examines the expressions for Hamiltonians, 
in particular, those terms which contain the mixing of two phonon modes. 
The mixing is contained mostly in the ``local'' term with $\beta$,
$$
\beta Q_2 \exp{i \pi( b_1^+ b_1)}.$$

The transfer term $T$ also contains the mixing, but it always enters the
expression
in the form $T(1+E_1E_2)$, thus for the transfer term its contribution 
is always shadowed by the larger term $\sim 1$. Thus, if $T\neq 0$ and
 large enough, its contribution is always dominant over mixing term $\beta$.
But if $T=0$, we are left with the only nonlinear coupling term
$\beta Q_2\varepsilon$ and accurate accounting for mode mixing becomes crucial.


The variational mean value of the  part of the 
Hamiltonian (\ref{diagh1}) in the state
(\ref{6}) renormalized by $\Omega$ yields (Appendix A)

\begin{eqnarray}
& & \frac{\langle H\rangle}{\Omega}=\frac{1}{2} (\cosh 4r_1+\cosh 4r_2)
\cosh 2\lambda +(\zeta_1^2+\zeta_2^2) \nonumber\\
& & + \frac{2\eta}{C}\frac{\exp
\left [-\frac{2\tilde\zeta_1^2}{\cosh 2\lambda}\right ]}{\cosh 2\lambda}
\left\{ -\tanh 2\lambda\sinh 2\lambda \cosh 2(r_1+r_2) \cosh
2(r_1-r_2)\right.\nonumber\\
& & \left.+\tilde\zeta_1^2[(\exp 2(r_1+r_2)-\exp(-2(r_1+r_2))\cosh 4\lambda)
(1+\tanh^2 2\lambda)\right.\nonumber\\
& & \left.+2 \exp(-2(r_1+r_2))\sinh 4\lambda\tanh 2\lambda]\cosh
2(r_1-r_2)\right.\nonumber\\
& & \left.+\frac{\tilde\zeta_1^2}{\cosh^2 2\lambda}(\sinh 4r_1-\sinh 4
r_2) -2\zeta_1(\tanh (2\lambda)
e^{2(r_2-r_1)}\zeta_2+\zeta_1)\right\} \\
& & +\frac{2\alpha}{\Omega C} (1-\eta^2) \zeta_1
\nonumber \\
& & -\frac{2\beta}{\Omega C} \left\{ (1+\eta^2)[\zeta_2 -\zeta_1\tanh 2\lambda \exp 2
(r_2-r_1)]\frac{\exp \left (-\frac{2\tilde \zeta_1^2}{\cosh 2\lambda
}\right )}{\cosh 2\lambda}
+2 \eta \zeta_2 \right\} \nonumber \, ,
\label{11}
\end{eqnarray}
where
\begin{equation}
\tilde\zeta_1= \zeta_1\exp(-2r_1), 
\quad C=1+2\eta \frac{\exp( -\frac{2\tilde\zeta_1^2}{\cosh2\lambda})}{\cosh
2\lambda}+\eta^2.
\label{12}
\end{equation}

From Eq. (\ref{11}) one can see that $\lambda$ causes correlations of the
selftrapping and tunneling dominated regions.

For $\lambda=0$, Eq. (\ref{11}) turns to
\begin{eqnarray}
& & \frac{\langle H\rangle_{\lambda=0} }{\Omega}= \frac{1}{2}(\cosh 4r_1+\cosh 4 r_2)+
\frac{1}{C_0}\left(1-2\eta e^{-8 r_1}\exp
(-2\tilde\zeta_1^2)+\eta^2\right)\zeta_1^2 \nonumber \\ & & +  \Omega\zeta_2^2 
 + \frac{2\alpha}{\Omega C_0}(1-\eta^2)\zeta_1- \frac{2\beta}{\Omega C_0}
 [(1+\eta^2)\exp (-2\tilde\zeta_1^2)+2\eta]\zeta_2
\label{13}
\end{eqnarray}
where
\begin{equation}
C_0 =1+2\eta\exp(-2\tilde \zeta_1^2)+\eta^2.
\label{14}
\end{equation}
For $\eta=0$, $ C=1$ and (\ref{11}) becomes
\begin{eqnarray}
& & \frac{\langle H\rangle_{\eta=0}}{\Omega}= \frac{1}{2}(\cosh 4r_1 +\cosh 4r_2)
\cosh 2\lambda + \zeta_1^2+\zeta_2^2 + \frac{2\alpha}{\Omega}\zeta_1\nonumber\\
& & -\frac{2\beta}{\Omega}(\zeta_2-\zeta_1\tanh 2\lambda\exp 2(r_2-r_1))\frac{\exp
\left (-\frac{2\tilde\zeta_1^2}{\cosh 2\lambda}\right )}{\cosh 2\lambda}
\label{15}
\end{eqnarray}
The ground states of (\ref{11}), (\ref{13}) and (\ref{15}) were found
by minimalization of the expressions against the involved VPs. The
respective ground state energies $E^{opt}$, $E_{\lambda=0}^{opt}$,
$E_{\eta=0}^{opt}$ and
$E_{\eta=0,\lambda=0, opt}$ will be compared mutually and with the exact
value from numerical simulation in order to find out
importance of the variational parameters $\eta, \lambda$ in different
regions of model parameters $\chi=\frac{\beta}{\alpha}$ (reflection) and
$\mu=\frac{\alpha^2}{\Omega^2}$ (effective e-ph coupling).
The place of E$\otimes$e JT model will clearly come out as an
important special case.

\section{Discussion of the numerical results}

In lattice electron-phonon models the nonadiabaticity parameter is the
ratio of phonon frequency and the band width, $\frac{\Omega}{T}$. The
energy is scaled by $T$ and the effective coupling parameter (ratio of
the polaron energy and of the phonon frequency) is $\mu_T= \alpha^2 /
2\Omega T$). In the present local model, the nonadiabaticity parameter
is the ratio of the frequency and the coupling parameter $\Omega/ \alpha$.
 The energy was scaled by $\Omega$ and the ratio of polaron energy and the
frequency is then $\alpha^2/2\Omega^2$. Therefore, the reduction of the
ground state energy of the local model due to selflocalization is stronger
than that of the lattice model by the factor $\mu_L/\mu_T= T/\Omega >1$.
The energy of the tunneling $\beta^2/\Omega^2$,
is comparable with the polaron energy at $\beta^2/\Omega^2=\mu_L$.
Because close to $\chi=1$ there is
$\zeta_1\approx 0$,  the effects of all variational parameters for
$\chi>1$ except of $\zeta_2$ vanish.
Quantum fluctuations due to $r_1, r_2$ and anharmonicity due to $\lambda$
are concentrated in the crossing region close to $\mu=1$ and $\chi=1$.

 \begin{figure}[hb]
\epsfxsize=11cm
\epsfbox{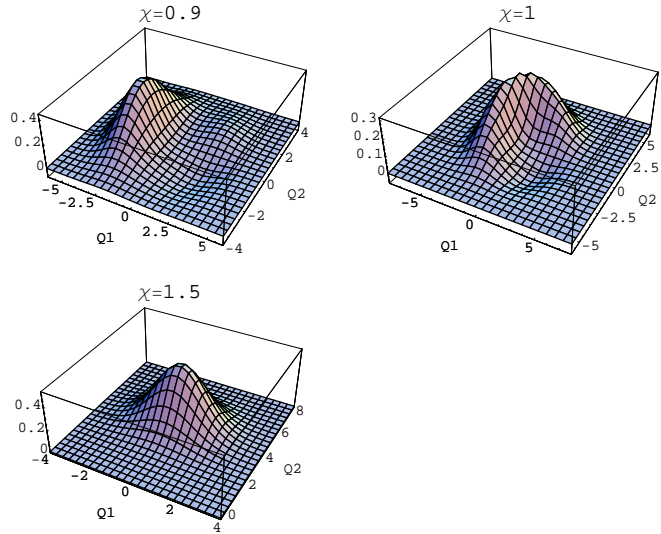}
\caption{The numerical ground state wave functions at $\mu= 2$ and
$\chi=0.9$ (a) $\chi=1$ (b) and $\chi=1.5$ (c).}
\label{fig4}
\end{figure}

The ground state in the phase plane $\chi $ and $\mu$ shown in
Fig.4 exhibits two phases separated by the crossover line close to $\chi=1$.
It means, that the effective polaron potential exhibits two minima which
reflect positions of two levels governed by the model parameters $\mu $
and $\chi$. The minima coincide within the border of the regions close to
$\chi=1$.
The phase $\chi<1$ is $\chi$ independent, selftrapping dominated,
with quantum fluctuations due to $\lambda, \eta, r_1, r_2 $. The phase
$\chi>1$ is the phonon-2-assisted tunneling dominated region with continuum
emission and absorption of virtual phonons-$1$. The phonon exchange
couples the levels within one minimum.
The minimum is much more sensitive to the
change of model parameters $\mu, \chi $ as well as to quantum fluctuations $r_1, r_2 $
and $\lambda$.

Electron in the selftrapping dominated region is trapped by the phonons-$1$ 
but due to the interactions mediated by phonons-2 the electron can
fluctuate to the higher level. Due to the reflection symmetry of the
phonons-2 continuum oscillations of the electrons at simultaneous
emission and absorption of phonons-1 occurs. These oscillations couple
the levels and so the electrons into pairs. This mechanism was described
in a recent paper \cite{Majernikova:2002} for a lattice model.

From (\ref{15}), for small $\chi $, one can approximately take
$\zeta_2\approx 0$, $\zeta_1\approx -\frac{\alpha}{\Omega}$ and for
$\lambda $ we are left with the result $\lambda\approx \mu\chi
\exp(-2\mu)$. This dependence can be recognized in Fig.5.

 \begin{figure}[h]
\epsfxsize=9cm
\epsfbox{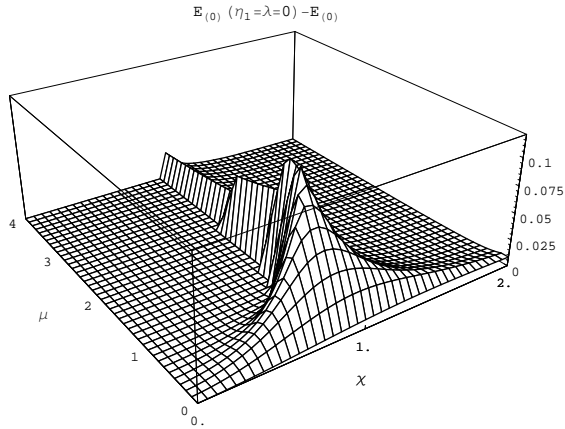}
\caption{The difference of $E(\eta=0,\lambda_{opt})-E(\eta_{opt},\lambda_{opt})$.}
\label{fig5}
\end{figure}

 \begin{figure}[h]
\epsfxsize=12cm
\epsfbox{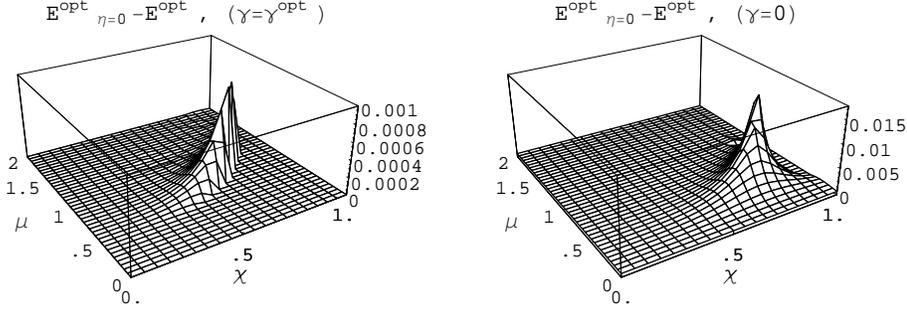}
\caption{The difference of $E(\eta=0,
\lambda_{opt})-E(\eta_{opt},\lambda_{opt})$ including the mode correlation
(a), without the correlation (b). Strong suppression of the reflection
effects by the mode correlation at $\mu<1$ is evident. The correlations
flatten the barrier between the zones and the role of the
parameter $\eta$ is minorized (Note the difference of the scales of
Figs. (a) and (b)).}
\label{fig6}
\end{figure}

Comparing Figs. 6a and 6b one can see that the mode correlation induced
anharmonicity  $\lambda$ (Fig.6a) is by one order larger than the contribution
to the anharmonicity of the reflection level mixing $\eta$ Fig.6b.
The correlation is most effective for weak effective couplings $\mu$ at
$\chi\approx 1$.
For large $\mu$ it contributes only close to $\chi=1$, where it reveals
maximum for all $\mu$. 
In order to check the validity of the calculations using variational
approach we performed also numerical
diagonalization of the Hamiltonian in the phonon-$1,2$ space.
We limit ourselves in $N_1$ phonon-1 states
and $N_2$ phonon-2 states, thus state vector is $N_1\times N_2$
dimensional. As numerical diagonalization results show,
about 20-50 phonon states are sufficient for convergence.
We show the results of numerical diagonalization of Hamiltonian
matrix as function of $\chi$ for $\mu=1$ and $\mu=4$. In the first
case we took 20$\times$20 state vector, while in the latter case to
achieve satisfactory convergence
(especially for the tunneling-dominated region when $\chi>1$)
we had to increase the number of phonons-2 up to 50.

\section{Quantum fluctuations in the local model}

Fluctuations of phonon-1 coordinate $\Delta Q_1^2=\langle Q_1^2\rangle -
\langle Q_1\rangle ^2$, $Q_1 = \frac{1}{\sqrt 2} (b_1^{\dag}+b_1)$ and
the conjugate momentum $P_1= \frac{-i}{\sqrt 2} (b_1^{\dag}-b_1)$ for
$\eta=0$ can be easily calculated analytically. The wave function of the
phonon-$1$ 
has the Gaussian form
$$
\phi(x) \sim \exp
\left[-\frac{(x-\lambda_1)^2}{\cosh(2\lambda)\exp(4r_1)}\right ]. $$
In fact, it is squeezed displaced harmonic oscillator
if one does not take into account the parameter $\eta$; Introducing
 $\lambda$ does not invoke higher order nonlinearities with respect to
phonon-$1$ coordinates. Thus $\langle\Delta P_1\rangle=0$, like it should be for an
harmonic oscillator, and the expressions for second momenta have the
form:

\begin{equation}
\Delta Q_1^2 = \frac{1}{2}\exp 4r_1\cosh 2\lambda, \quad
\Delta P_1^2 = \frac{1}{2}\exp(- 4r_1)\cosh 2\lambda
\label{20a}
\end{equation}
and
\begin{equation}
\left (\Delta P_1^2\Delta Q_1^2\right )^{1/2}= \frac{1}{2}\cosh
2\lambda.
\label{21}
\end{equation}
The shape of fluctuations due to the correlation $\lambda$ (\ref{21})
closely follows the shape of the curve in Fig.7, having its maximum
close to $ \chi=1$ and $\mu \approx 1$.

 \begin{figure}[h]
\epsfxsize=10cm
\epsfbox{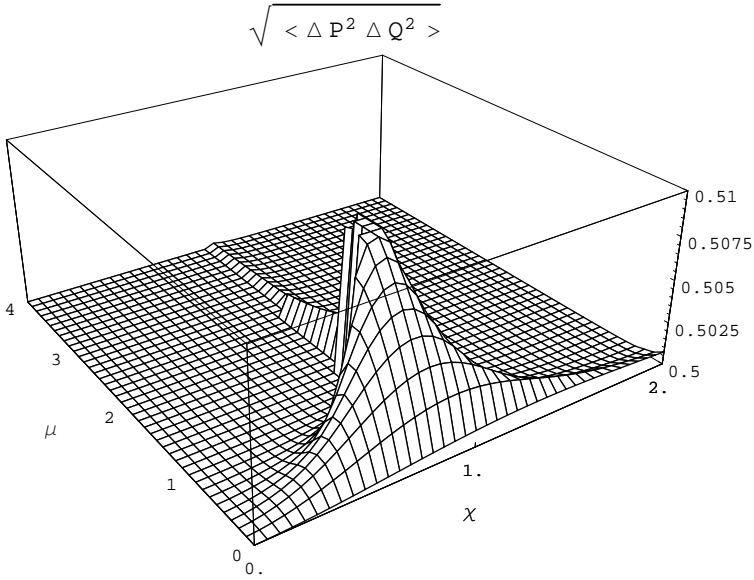}
\caption{The product of uncertainties of the coordinate and the
conjugated momentum of the phonon-$1$ for $\eta=0$. }
\label{fig7}
\end{figure}

The similarity of the product of uncertainties of
phonon coordinate and momentum displayed in Fig.7. illustrates anharmonic
source of the contribution due to the mode correlation.
 The correlations $\lambda $ manifest themselves in (\ref{20a}) and
(\ref{21}) as a phonon anharmonicity. It is noteworthy to emphasize its
quantum origin (compare the note below Eq. (\ref{1})).
The problem with $\eta\neq 0$ needs special consideration and will be
considered elsewhere.

\section{Conclusion}

   A suitable choice of the variational wave functions for various
electron-phonon two-level systems is a long-standing problem in solid state
physics as well as in quantum optics.
For two-level reflection symmetric systems with intralevel electron-phonon interaction
 the approach with a variational two-center
squeezed coherent phonon wave function was found to yield the lowest
ground state energy. The two-center wave function was constructed as a
linear combination of the phonon wave functions related to both levels
introducing new VP.

This symmetry implies coupling of the levels mediated by continuous
virtual emission and absorption of
phonons accompanying the electron tunneling between the levels.
This is an analogy of Rabi oscillations in quantum optics.
Investigation of referring properties (tunneling rates,
 Debye-Waller factors) was performed mostly for the models
with onsite (Holstein) electron-phonon interactions in dimer
or exciton-phonon models and also for E$\otimes$e JT model. The methods
used there were based on either combination of variational approach and unitary
transformations or numerical diagonalization.

In two-level models with phonon assisted tunneling there appears
coupling of both phonon modes mediated by Rabi oscillations. Therefore, also
the interlevel quantum  correlations of the phonons must be taken into account.
This brings into the variational ansatz additional quantum VP.


The two-level system is characterized by strongly nonlinear
effective potential provided by phonons.
The nonlinearity is governed by the bare Hamiltonian
parameters exhibiting a crossover from an asymmetric double
potential well  with two broad nonequivalent minima pertaining to the
levels to a regime of one
narrow minimum when the broad minima coincide into one potential well.
This effect is due to the effective attraction of the levels by virtual
exchange of phonons.
Much effort was expended also to improve one-center  variational approach
by including correlations of the concerning two squeezed coherent phonon
modes of the JT model. When compared to the lattice case,
the local version of the model provides
 qualitatively similar effective potential.
This conclusion was to be
expected: in the symmetric case minima of the nonlinear effective
potential related with two levels are close together  and so the quantum
fluctuations there are most effective.
We considered both the mixing and squeezing
correlation VP in order to find region of relevance of both parameters as
functions of the interaction constants.\\

The support from the Grant Agency of the Czech Republic of our project
No. 202/01/1450 is highly acknowledged. We thank also the grant agency VEGA (No.
2/7174/20) for partial support.

\vspace{0.5cm}
{\bf Appendix A}

We used following formulas \cite{Bonny:1986}
\begin{eqnarray}
& & D_{1}(\eta_1)S_1(r_1)= S_{1}(r_1)D_{1}(\tilde \eta_1), \
\tilde\eta_1=\eta_1 e^{-2r}
\label{20}\\
& & S_1^{-1}(r_1)b_1 S_1(r_1) = b_1\cosh 2r_1+ b_1^{\dag} \sinh 2r_1
\label{211} \\
& & S_{12}^{-1}(\lambda) b_1 S_{12}(\lambda)= b_1\cosh \lambda + b_2^{\dag}
\sinh \lambda
\label{22} \\
& & \langle 0|S_1^{\dag}(r)D_{1}^{\dag}(\eta)\exp(\lambda
b_1^{\dag})\rangle= \frac{1}{(\cosh 2r)^{1/2}}\nonumber\\
& & \times\exp \left [\frac{\lambda^2}{2}\tanh (2r) -\lambda\eta
 (\tanh 2r -1)-
\frac{1}{2}\eta^2+\frac{1}{2} (\tanh 2r -1)\eta^2 \right ]
\label{23}, \\
& & \langle 0|S_1(\zeta)D_{1}(\eta)b_1^{\dag m}
\rangle=\frac{d^m}{d\lambda^m}
\langle 0|S_1(\zeta)D_{1}(\eta )\exp(\lambda
b_1^{\dag})\rangle  |_{\lambda=0},
\label{24}\\
& &  S_{12}(\lambda)= T^{\dag}\left (\frac{\pi}{4} 
\right )S_1 (\lambda/2) S_2 (-\lambda/2
)T\left (\frac{\pi}{4} \right )
\label{25} \\
& & T\left(\frac{\pi}{4}\right )\left (\matrix {b_1\cr b_2}\right )
T^{\dag}\left(\frac{\pi}{4}\right )=
\frac{1}{\sqrt 2} \left (\matrix {b_1-b_2\cr b_1+b_2}\right ),
\label{26} \\
& & T^{\dag}|0\rangle =T|0\rangle =|0\rangle
\label{27}
\end{eqnarray}


\begin{thebibliography}{10}
\bibitem{Obrien:1993} M.\ C.\ M. O'Brien and C.\ C. Chancey,   
Am.J.Phys. {61} (8), 688 (1993)
\bibitem{Kaplan:1995} M.\ D. Kaplan and B.\ G. Vekhter,  ``Cooperative
phenomena in Jahn-Teller crystals'' (Ed. J.\ P.\ Fackler, Plenum, New York 
and London 1995).
\bibitem{Gunnarson:1995} O. Gunnarson, Phys.\ Rev.\ Lett. { 74}, 1875
(1995); O. Gunnarson, Rev.\ Mod.\ Phys. { 69}, 575 (1997).
\bibitem{Sander:1973} H.\ B.\ Shore and L.\ M.\ Sander, Phys.\ Rev.\ B{ 7} (10),
4537 (1973)
\bibitem{Wagner:1994} M. Sonnek, T. Frank and M. Wagner, Phys.\ Rev.\ B
{ 49} (22), 15637 (1994)
\bibitem{Lo:1991a} C.F. Lo, Phys.\ Rev.\ A { 43} (9), 5127 (1991)
\bibitem{Lo:1991b} C.F. Lo and R. Sollie, Phys.Rev.\ B { 44} (10), 5013 (1991)
\bibitem{Fulton:1961} R.L. Fulton and M. Gouterman, J.\ Chem.\ Phys.
{35}, 1059 (1961).
\bibitem{Rabi:1936} B. W. Shore, and P.\ L. Knight, J.\ Mod.\ Opt. { 40},
1195 (1993); I.I. Rabi, Phys.Rev. { 49}, 324 (1936); Phys.\ Rev. {
51}, 652 (1937).
\bibitem{Majernikova:2002} E. Majern\'{\i}kov\'a, J. Riedel and S. Shpyrko,
Phys.\ Rev.\ B { 65} (17), 174305 (2002)
\bibitem{Bonny:1986} B. L. Schumaker, Phys.\ Repts. {133} (6), 317
(1986)
\end{thebibliography}
\end{document}